\documentclass[]{aastex631}

\begin{document}

\title{A recent re-acceleration of the Local Bubble revealed by kinematics of young star associations\footnote{}}

\correspondingauthor{Guang-Xing Li}
\email{gxli@ynu.edu.cn}

\correspondingauthor{Bing-Qiu Chen}
\email{bchen@ynu.edu.cn}

\correspondingauthor{Krause G.H. Martin}
\email{m.g.h.krause@herts.ac.uk}

\author{Guang-Ya Zeng}
\affiliation{South-Western Institute for Astronomy Research, Yunnan University, Chenggong District, Kunming 650500, P. R. China}

%\collaboration{20}{(AAS Journals Data Editors)}

\author{Guang-Xing Li}
\affiliation{South-Western Institute for Astronomy Research, Yunnan University, Chenggong District, Kunming 650500, P. R. China}

\author[0000-0003-2472-4903]{Bing-Qiu Chen}
\affiliation{South-Western Institute for Astronomy Research, Yunnan University, Chenggong District, Kunming 650500, P. R. China}

\author{Ji-Xuan Zhou}
\affiliation{School of Physics and Astronomy, Cardiff University, Queen’ s Buildings, The Parade, Cardiff CF24 3AA, UK}

\author[0000-0002-9610-5629]{Martin G.H. Krause}
\affiliation{Centre for Astrophysics Research, School of Physics, Astronomy and Mathematics, University of
Hertfordshire, College Lane, Hatfield, Hertfordshire AL10 9AB, UK}

%% Note that the \and command from previous versions of AASTeX is now
%% depreciated in this version as it is no longer necessary. AASTeX 
%% automatically takes care of all commas and "and"s between authors names.

%% AASTeX 6.31 has the new \collaboration and \nocollaboration commands to
%% provide the collaboration status of a group of authors. These commands 
%% can be used either before or after the list of corresponding authors. The
%% argument for \collaboration is the collaboration identifier. Authors are
%% encouraged to surround collaboration identifiers with ()s. The 
%% \nocollaboration command takes no argument and exists to indicate that
%% the nearby authors are not part of surrounding collaborations.

%% Mark off the abstract in the ``abstract'' environment. 
\begin{abstract}

The low-density region of the interstellar medium (ISM) where the Sun is located is known as the Local Bubble, a cavity filled with high-temperature and low-density plasma that may be created by a series of supernova (SN) explosions over the past 14 Myr. However, the effects of these SN explosions on the formation and evolution of the Local Bubble, as well as on nearby star formation, remain not fully understood. To study the expansion history of the Local Bubble, we use the kinematic data of the young stars obtained by cross-matching the pre-main-sequence (PMS) star catalog of \citet{Zari2018} with the high-precision astrometric and photometric data from the {\it Gaia} DR3 database. We perform a three-dimensional spatial clustering analysis on these young stars to identify star associations. We discover three unique star associations that exhibit a wiggle-like velocity pattern. The distances of these star associations are 108.5308, 141.5284, and 176.0318 pc, respectively. Their radial velocities in the Local Standard of Rest (LSR) are 10.0622, 5.4982, and 9.0581 km/s, showing a pattern of decreasing and then increasing. This velocity pattern, as predicted by \citet{Krause&Diehl2014}, is caused by a recent re-acceleration affected by the SN explosion, reinforcing the picture of the Local Bubble as an evolving entity.

\end{abstract}

%% Keywords should appear after the \end{abstract} command. 
%% The AAS Journals now uses Unified Astronomy Thesaurus concepts:
%% https://astrothesaurus.org
%% You will be asked to selected these concepts during the submission process
%% but this old "keyword" functionality is maintained in case authors want
%% to include these concepts in their preprints.
\keywords{Local Bubble --- Pre-main-sequence stars --- Kinematics}

%% From the front matter, we move on to the body of the paper.
%% Sections are demarcated by \section and \subsection, respectively.
%% Observe the use of the LaTeX \label
%% command after the \subsection to give a symbolic KEY to the
%% subsection for cross-referencing in a \ref command.
%% You can use LaTeX's \ref and \label commands to keep track of
%% cross-references to sections, equations, tables, and figures.
%% That way, if you change the order of any elements, LaTeX will
%% automatically renumber them.
%%
%% We recommend that authors also use the natbib \citep
%% and \citet commands to identify citations.  The citations are
%% tied to the reference list via symbolic KEYs. The KEY corresponds
%% to the KEY in the \bibitem in the reference list below. 

\section{Introduction} \label{sec:intro}

%\latex\ \footnote{\url{http://www.latex-project.org/}}

Stars directly impact their surrounding ISM through the emission of ionizing radiation, stellar winds and SN explosions (at the end of the most massive stars' lives). These processes are collectively termed stellar feedback \citep{Collins&Read2022}. Stellar feedback stands as one of the key drivers of galaxy evolution. It profoundly shapes the structure of ISM, as SN explosion serve as a key source of momentum, energy, and mass reintroduced into the ISM \citep{Porter2024}.

Stars often form in cluster \citep[e.g.,][]{Larsen1999,Krause2020}. Thus, bubbles produced by individual massive stars frequently combine into superbubbles. The term is often used in the X-ray community \citep[e.g.,][]{Dunne2001}, where the superbubble becomes X-ray-bright for some time after each SN explosion inside the superbubble \citep{Krause2014,Krause&Diehl2014}. While delivering feedback energy into a common superbubble leads to less radiative cooling compared to individual isolated bubbles, 3D instabilities of the shell lead overall to rapid energy loss \citep{Krause2013,Lancaster2021,Lancaster2024}. Supershells therefore expand most of the time at moderate velocities, with short-term accelerations when massive stars increase their energy output, e.g., in a SN explosion. This means that stars formed in or triggered by the shell may after several Myr have a variety of velocities, and are generally expected to be found inside the superbubble. Note that this is different to outdated 1D superbubble models with continuous energy injection \citep[e.g.,][]{Weaver1977}, where the supershell continuously decelerates, and thus any stars formed in the shell would keep their formation velocity and outrun the supershell.

The low-density region of ISM where the Sun is located is known as the Local Bubble, filled with low-density and high-temperature plasma detectable by soft X-rays \citep[e.g.,][]{Cox&Reynolds1987,Liu2017,Zheng2024}. This cavity, which surrounds the Sun, is 100 to 200 pc wide and is surrounded by a layer of cold, neutral gas and dust \citep{Zucker2022}. Measurements of Na I and Ca II absorption lines suggest that the Local Bubble lacks neutral gas \citep{Welsh2010}, and the observations in the soft X-ray have verified that the internal temperature of the Local Bubble reaches millions of Kelvin \citep{Snowden1997}. The conclusive interpretation of all these characteristics is that the Local Bubble was formed by the stellar winds of nearby massive stars and a series of SN explosions over the past 10 Myr \citep{Smith&Cox2001,Schulreich2023}.

\citet{Ratzenbock2023} reconstructed the star formation history of the nearest OB association to Earth, Sco-Cen, by deriving the ages of the 37 clusters selected by the \textbf{SigMA} algorithm. They identified four distinct phases of enhanced star formation activity, which roughly occurred during bursts at 20 Myr, 15 Myr, 10 Myr, and 5 Myr ago. The star formation history of Sco-Cen is dominated by a brief period of star and cluster formation rate at 15 Myr ago. This burst is consistent with the scenario proposed by \citet{Zucker2022}, which suggests that the Local Bubble was triggered by massive stellar feedback from Sco-Cen.

\citet{Fuchs2006} found that young clusters entered the current Local Bubble region around 10 $\sim$ 15 Myr ago and estimated 14 $\sim$ 20 massive members have exploded since that time, through kinematic analysis of the solar neighborhood within 200 pc. \citet{Breitschwerdt&deAvillez2006} have constrained the age of the Local Bubble, according to 19 SN explosions have occurred to date, to $14.5^{+0.7}_{-0.4}$ Myr through 3D high-resolution hydrodynamic simulations. \citet{Zucker2022} also found that the majority of nearby star-forming regions are positioned on the surface of the Local Bubble, indicating that the expansion of the Local Bubble is responsible for triggering nearly all star formation in the vicinity. However, there are still many uncertainties regarding the formation and evolution of the Local Bubble, the process of SN explosion, and their impact on nearby young stars and star formation.

\citet{Swiggum2025} performed backward orbital integrations for 509 clusters within 1 kpc of the Sun, tracing their evolution over the past 100 Myr. The majority of the young clusters are divided into three spatial groups, which trace the Pleiades, Coma Berenices, and Sirius moving groups, respectively. They suggest that these young clusters likely formed in massive star-forming complexes, initially moving together with the older stars in the moving groups before gradually separating after the parent gas was expelled by stellar feedback. The kinematic features inherited by the young clusters are likely formed under the influence of Galactic-scale perturbations driven by spiral arms.

Young stars typically form within vast molecular clouds, where they are accompanied by hundreds or thousands of other young stars \citep{Kounkel2022}. It is difficult to directly detect the three-dimensional spatial motion of the molecular gas, but since young stars are unlikely to deviate far from their natal clouds and their kinematic information is similar to that of their natal molecular clouds \citep{Tu2022}, it is feasible to track the motion of these molecular gases based on the kinematic data of young stars born within them. At the same time, as age increases, the "initial" structural characteristics tend to be blurry as a result of n-body interactions among stars. Therefore, the structural and kinematic characteristics associated with molecular clouds are likely to be more pronounced in young star clusters \citep{Krause2020}.

We study the kinematic characteristics of star associations near the Local Bubble by combining the pre-main-sequence stars in the PMS catalog of \citet{Zari2018} with the high-precision photometry and kinematic data provided in the {\it Gaia} DR3 catalog. The organization of this paper is as follows. We describe the data sample used in Section \ref{sec:data}, present the methods for obtaining the kinematic characteristics of star associations in Section \ref{sec:CKP}, discuss and analyze the results of our work in Section \ref{sec:SSA} and \ref{sec:RSWVP}, and provide a summary in Section \ref{sec:summary}.

\section{Data} \label{sec:data}

The young star sample we use comes from the PMS catalog of \citet{Zari2018}, which includes a total of 43,719 sources. They utilized data from {\it Gaia} DR2, selecting stars within a 500 pc neighborhood of the Sun. By applying comprehensive constraints on photometric and astrometric parameters, they filtered out early-type stars and pre-main-sequence stars on the Hertzsprung-Russell diagram, and the sample's extinction correction was made using a three-dimensional g-band extinction map produced by the Zari’s themselves.

Compared to {\it Gaia} DR2, {\it Gaia} DR3 \citep{Gaia_Collaboration_Vallenari2022} is based on astrometric and photometric measurements of over 1.8 billion celestial bodies observed over 34 months, providing higher precision for astrometric parameters such as celestial coordinates, proper motions, parallaxes, as well as photometric parameters across various bands, and also includes a wealth of radial velocity information. We cross-match the PMS catalog of \citet{Zari2018} with {\it Gaia} DR3 to obtain more accurate data. We select all sources that match with the radial velocities from {\it Gaia} DR3, amounting to a total of 19,079, which serves as the sample for processing and analysis.

%\noindent {\tt\string\documentclass[twocolumn]\{aastex631\}}. \\

%{\tt\string\begin\{acknowledgments\}} ... {\tt\string\end\{acknowledgments\}}

\section{Clustering and Kinematic Parameters} \label{sec:CKP}

To study the kinematic characteristics of star associations and molecular clouds in the solar neighborhood, we perform a clustering analysis on the selected pre-main-sequence star samples to extract the structures of star associations and assign each pre-main-sequence star to the corresponding structure.

\begin{figure}[ht!]
\plotone{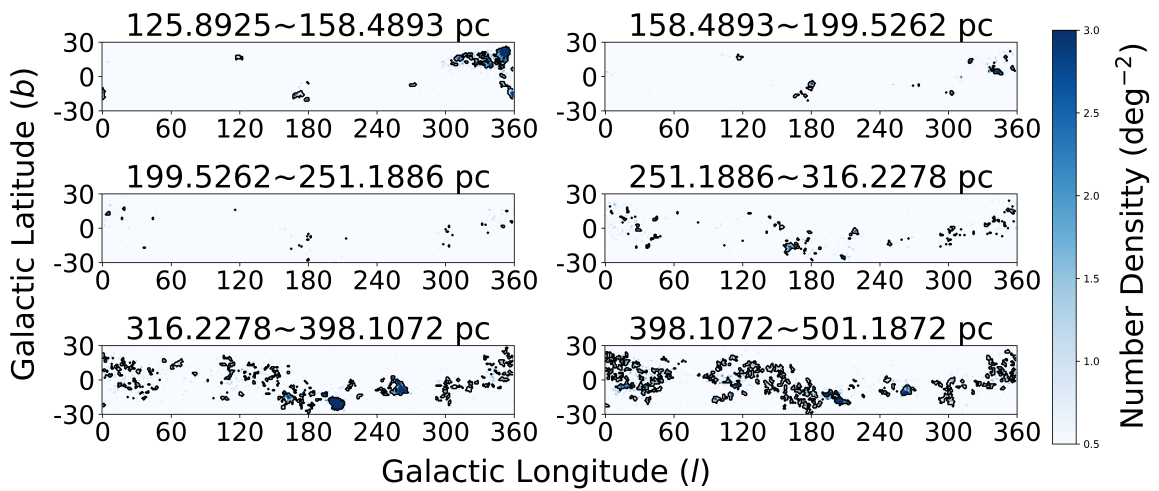}
\caption{The spatial density distribution of our pre-main sequence star sample (a total of 19,079 sources) across partial distance bins, along with the identified star associations. The color represents the number density of pre-main sequence stars, with the solid black line delineating the boundaries of the lowest-level substructures found by the \textbf{Dendrogram}. 
\label{fig:fig_1}}
\end{figure}

We use the \textbf{histogramdd} module in \textbf{python} to create a density map of our samples in the three-dimensional space of Galactic longitude, latitude, and distance. The ranges for each are $0^{\circ} < l < 360^{\circ}$, $-90^{\circ} < b < 90^{\circ}$, and $1.3 < {\rm log}_{10}(d/{\rm pc}) < 3.8$. For the density map, we apply a Gaussian filter for smoothing. Partial results are depicted in Figure \ref{fig:fig_1}.

For the smoothed density map, we utilize the \textbf{Dendrogram} program from the astrodendro module in Python to extract structures. It offers a straightforward method to compute dendrograms for observational or simulated astronomical data in Python. Like other dendrogram tools, it employs a hierarchical clustering algorithm. The program initially calculates the attribute distance between each pair of classes in the input feature file. It then iteratively merges the closest pair of classes, proceeding to the next closest pair after completion, until all classes have been merged. Ultimately, we identify 263 structures.

Due to the fact that both the method of screening pre-main-sequence stars and the method of extracting structures can introduce a significant number of contaminating sources, it is necessary to perform a filtering process on the sources assigned to each structure. We employ the method from \citet{Zhou2022}, which involves verifying the member stars of each structure through the distribution of distances. Ultimately, 180 star association structures remain.

To calculate the kinematic data such as the Galactic coordinates, distance, velocities in different directions, velocity dispersion, LSR\footnote{The LSR is a reference frame in astronomy, the motion of which is consistent with the average velocity of all stars within a local region near the Sun (typically within 100 parsecs or more from the Sun) within the Milky Way.} radial velocity, and peculiar velocities of the identified star associations, we employ various method. The Galactic coordinates, distance and velocities in different directions are determined using the average values and standard deviations of the parameters of the member stars within the star association. The velocity dispersion is calculated using the formula $\sigma_{v}  = \sqrt{{\rm std}(v_{i})^{2} - {\rm mean}(v_{{\rm error},i})^{2}}$. Here, $v_{i}$ and $v_{{\rm error},i}$ represent the velocities and velocity errors of the member stars in different directions, respectively.

To obtain the LSR radial velocity of the star associations, we calculate the LSR radial velocities of the member stars in the star associations based on the heliocentric radial velocities provided by {\it Gaia} DR3, using the formula from \citet{Reid2009}: $v_{r,\rm{LSR}} = v_{r,\rm{Helio}} + [U_{\odot}^{\rm{std}} \times {\rm cos}(l) +V_{\odot}^{\rm{std}} \times {\rm sin}(l) ] \times {\rm cos}(b) +  W_{\odot}^{\rm{std}} \times {\rm sin}(b)$. The parameters in the formula, $U_{\odot}^{\rm{std}} = 11.69$ km/s, $V_{\odot}^{\rm{std}} = 10.16$ km/s, and $W_{\odot}^{\rm{std}} = 7.69$ km/s, are derived from the results of \citet{Wang2021}. Then, we use Kernel Density Estimation\footnote{KDE is a common non-parametric estimation method in statistics, used to estimate the probability density function of a random variable. KDE employs kernel functions and, based on a certain bandwidth parameter, estimates the probability density of data points by taking a weighted average of the kernel functions around each data point. In other words, it infers the overall distribution from a finite sample of data.} (KDE) to fit these data and obtain the LSR radial velocity of the association. The kernel function we used is the Gaussian kernel.

We project the 3D-LSR velocities of the various star associations onto the dynamical LSR\footnote{The dynamical LSR is identical to the LSR, and its motion also follows the average motion of the stars. It is an inertial frame of reference that is transposed continuously along a circular orbit, and on which the acceleration directed toward the Galactic center is eliminated.} reference frame and the Local Co-rotating Frame\footnote{The LCF is a reference frame defined by \citet{Li2022}. Similar to the dynamical LSR, the LCF moves along a circular orbit around the Galactic center. It includes rotation around the Z-axis, which is added to ensure that the x-axis of the frame is locked toward the Galactic center. Due to this, LCF becomes non-inertial, and calculations performed in this frame should take into account the Coriolis force. In the LCF, the radial component of the velocity should be consistent with $v_{r,\rm{LSR}}$.} (LCF). In the LCF, the net rotation is removed to reveal shear. Subsequently, we subtract the assumed Galactic shear to reveal the peculiar velocities.

\section{Special Star Associations}\label{sec:SSA}

We discover three unique star associations (named Phantom, Shadow-cluster, and Bubble-edge from near to far) located within wall of the Local Bubble, including a gasless one. These star associations are spatially close and are the most prominent structures within the region defined by $-90^{\circ} < l < 30^{\circ}$, $-60^{\circ} < b < 60^{\circ}$, and $80\;{\rm pc} < d < 220\;{\rm pc}$ (Detailed in Appendix \ref{APP:B}). At the same time, they exhibit a wiggle-like velocity pattern. The distances of Phantom, Shadow-cluster, and Bubble-edge are 108.5305, 141.5284, and 176.0318 pc, respectively, while their LSR radial velocities are 10.0622, 5.4982, and 9.0581 km/s, showing a pattern of decreasing and then increasing.

\begin{figure}[ht!]
\plotone{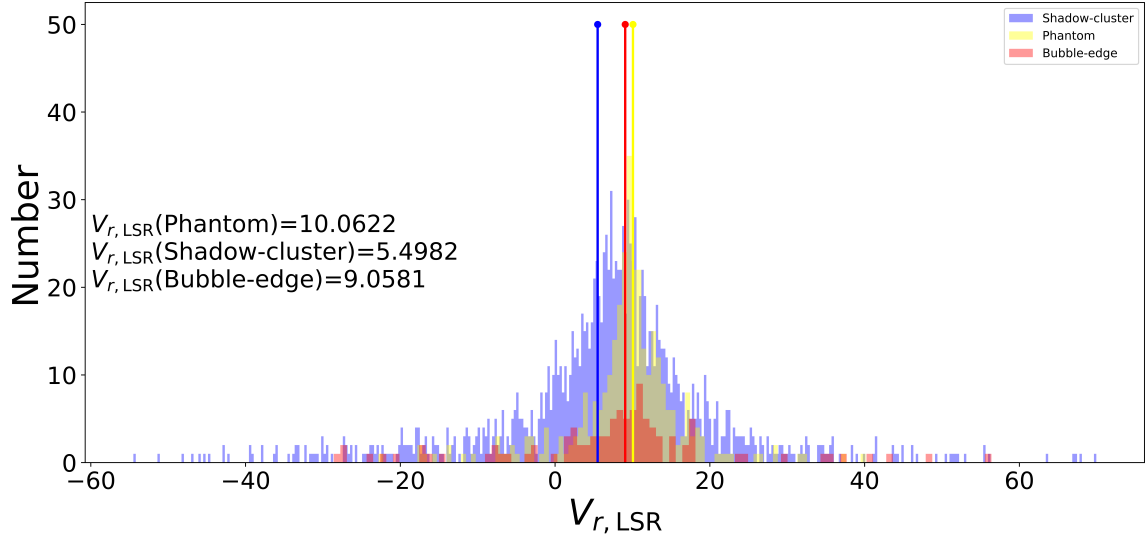}
\caption{Histogram of the LSR radial velocities of the star associations member stars calculated using parameter $U_{\odot}^{\rm{std}} = 11.69$ km/s, $V_{\odot}^{\rm{std}} = 10.16$ km/s, and $W_{\odot}^{\rm{std}} = 7.69$ km/s \citep{Wang2021}, with the vertical line (with a dot at the top) indicating the LSR radial velocity of the star association obtained through KDE fitting. 
\label{fig:fig_2}}
\end{figure}

\begin{figure}[ht!]
\plotone{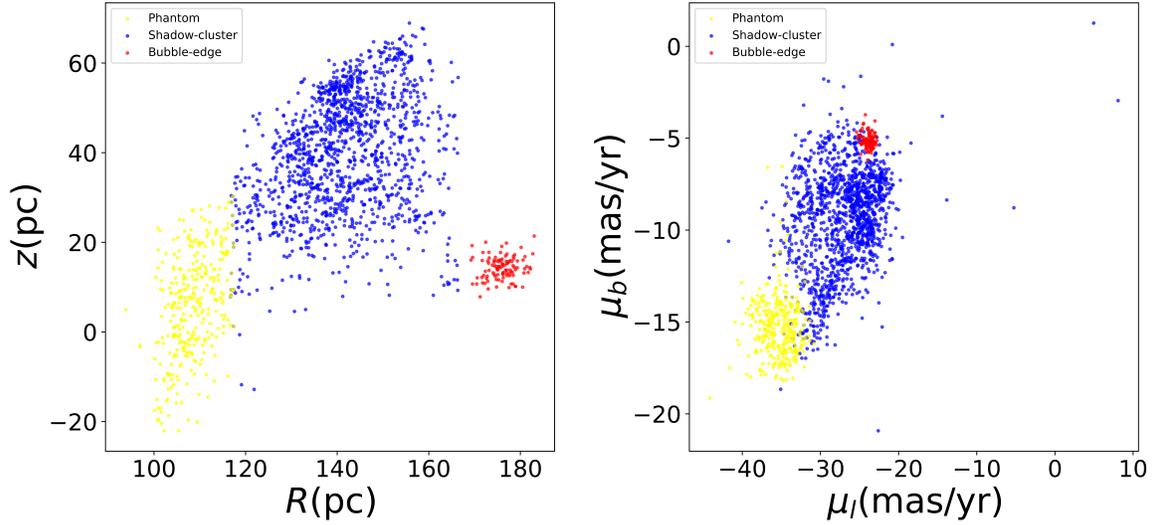}
\caption{Distribution of the special star associations in the R-z space and the $\mu_{l}$-$\mu_{b}$ space, with different colors indicating member stars of different star associations. 
\label{fig:fig_3}}
\end{figure}

Figure \ref{fig:fig_2} shows the distribution of LSR radial velocities for star associations Phantom, Shadow-cluster, and Bubble-edge. We note that all velocity centroids\footnote{The velocity centroid refers to the weighted average of the motion velocity of a system or object within a certain observed velocity space region. It is similar to the concept of the center of mass in space.} are positive, i.e., outwards, as expected if these stars were produced in the expanding supershell. The left side of Figure \ref{fig:fig_3} shows the distribution of these star associations in the distance(d)-height(z) space, from which it can be seen that Phantom, Shadow-cluster, and Bubble-edge are located at different distances, ranging from near to far, at 108.5308, 141.5284, and 176.0318 pc, respectively. We find that the LSR radial velocities of Phantom and Bubble-edge are higher than the typical expansion velocity of the Local Bubble \citep[6.7 km/s,][]{Zucker2022}. The specific reasons for this situation are explained in Appendix \ref{APP:A}.

\begin{figure}[ht!]
\plotone{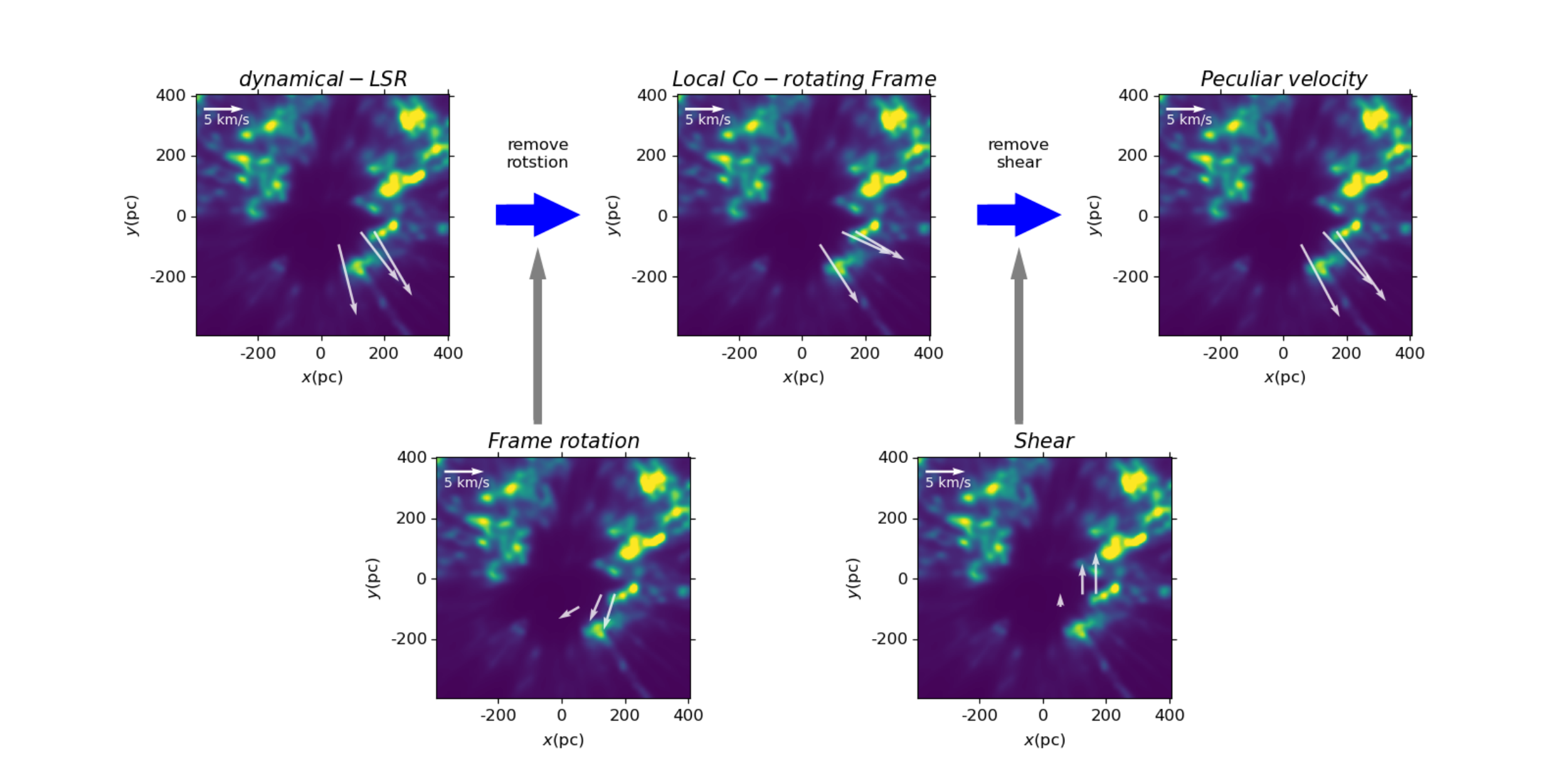}
\caption{This figure illustrates the process of velocity transformation in different reference frames for these unique star associations, including the dynamical LSR frame, Local Co-rotating frame, and peculiar velocities.
\label{fig:fig_4}}
\end{figure}

In Figure \ref{fig:fig_4}, we illustrate the process of velocity transformation in different reference frames, as described in Section \ref{sec:CKP}, using these unique star associations as examples.

\begin{figure}[ht!]
\plotone{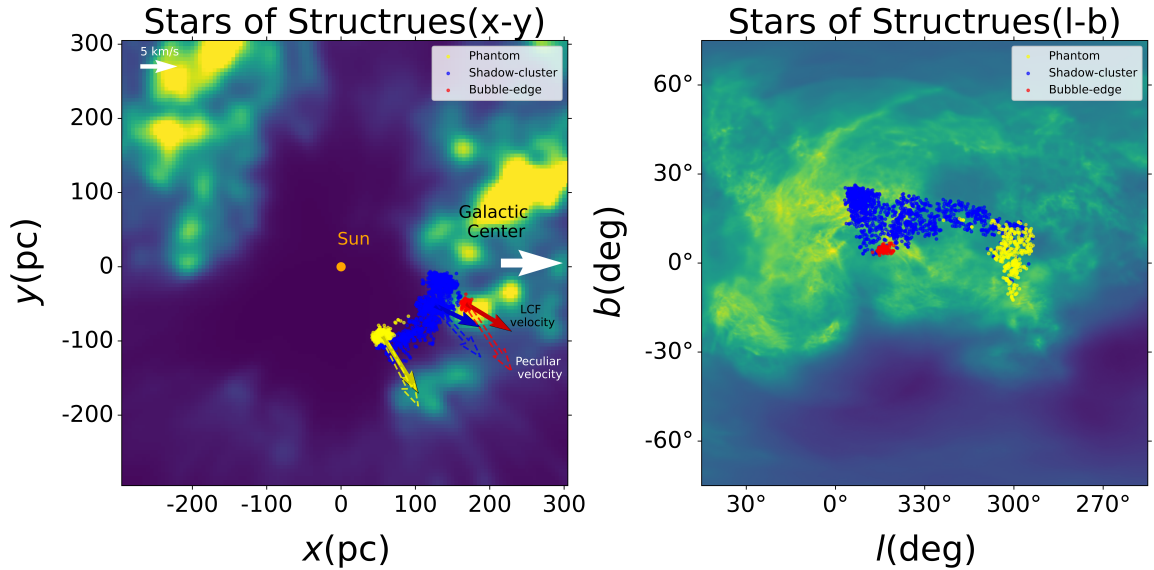}
\caption{Distribution of special star associations in the x-y and l-b spaces with different colors indicating member stars of different star associations. Arrows indicate the peculiar velocities (hollow arrows) and velocities in the LCF(solid arrows) of the star associations. The orange circle representing the Sun and the white arrow points towards the direction of the Galactic Center. The background of the left figure is the x-y plane of the 3D (x, y, z) dust image in the solar neighborhood and the background of the right figure is the l-b plane of the 3D (l, b, d) interstellar dust spatial distribution map drawn by \citet{Edenhofer2023}
\label{fig:fig_5}}
\end{figure}

Figure \ref{fig:fig_5} shows the positions of Phantom, Shadow-cluster, and Bubble-edge in the x-y space and their peculiar velocities, as well as their positions in the l-b space, with the right side of Figure \ref{fig:fig_3} showing the distribution of proper motion in the Galactic longitude and latitude directions. It can be seen that although Phantom, Shadow-cluster, and Bubble-edge are very close in space, they still have sufficient separability to be distinguished as different star associations. This property is also reflected in the distribution of proper motion. At the same time, their peculiar velocities and LSR radial velocities exhibit the nature of the Local Bubble's expansion. The elongated, curved nature of Shadow-cluster may be reminiscent of stars that formed in an expanding supershell.

We compared the kinematic characteristics of these three star associations with the moving groups studied by \citet{Swiggum2025}, but no one-to-one correspondence was found. At the same time, our sample also has a significant age gap compared to the clusters of \citet{Swiggum2025}. This suggests that the factors leading to the characteristic velocity distribution do not originate from within the molecular clouds, but are more likely to be caused by external influences.

\section{Recurrent Supernova and Wiggle-like Velocity Pattern} \label{sec:RSWVP}

\begin{figure}[ht!]
\plotone{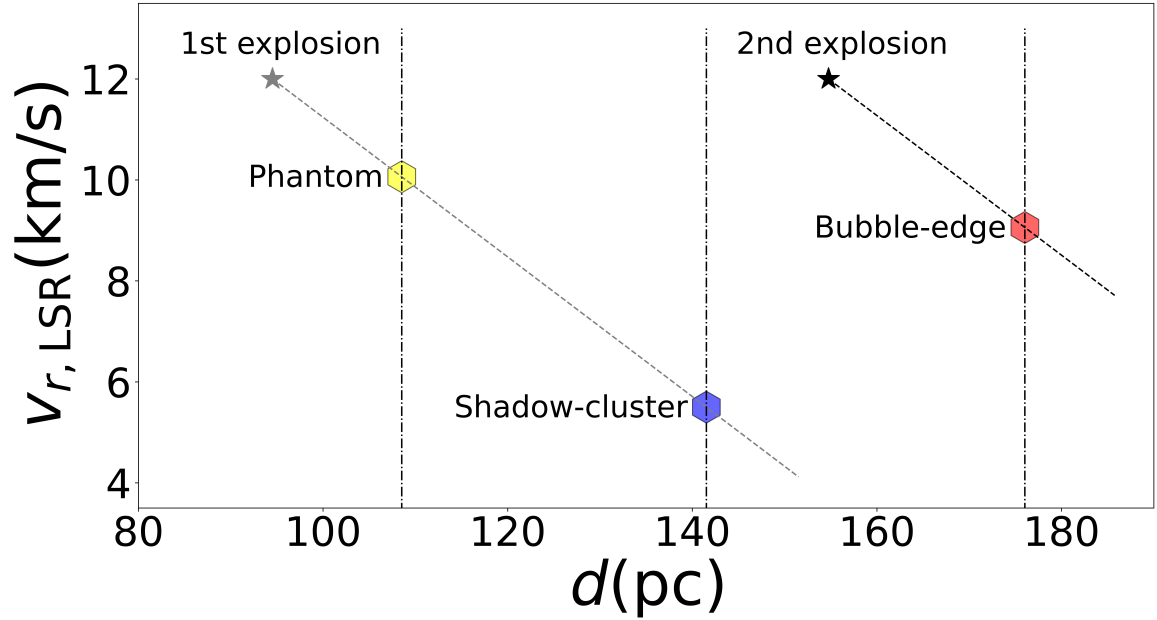}
\caption{An illustration of our hypothesis that different sub-associations within the star associations have been affected by multiple SN explosions. The hexagonal points in different colors represent the various star associations, the gray and black stars denote two SN explosions, and the dashed lines indicate the reduction in molecular cloud velocity after SN explosion and the process of star association formation affected by the SN explosion.
\label{fig:fig_6}}
\end{figure}

Detailed 3D hydrodynamics simulations of SuperBubble (SB) show that supershells decelerate most of the time and experience brief episodes of acceleration after each supernova \citep{Krause&Diehl2014}. We propose that these association may have been influenced by multiple SN explosions during the formation of the Local Bubble. Following the first SN explosion event, the molecular cloud expands outward under the drive of the ionization front, with its expansion speed gradually decreasing, and stars forming in this process. Due to the diminished influence of SN explosion after the formation of stars, the velocity of star associations stabilizes. Consequently, Phantom and Shadow-cluster formed successively and have essentially maintained the velocities they had at the time of their formation. Bubble-edge likely formed as a result of the second SN explosion event. At the time of star formation, the molecular cloud had a higher velocity. This ultimately led to these star associations being close together in space but separated in velocity, creating a wiggle-like velocity pattern. We illustrate this process in Figure \ref{fig:fig_6}.

\begin{figure}[ht!]
\plotone{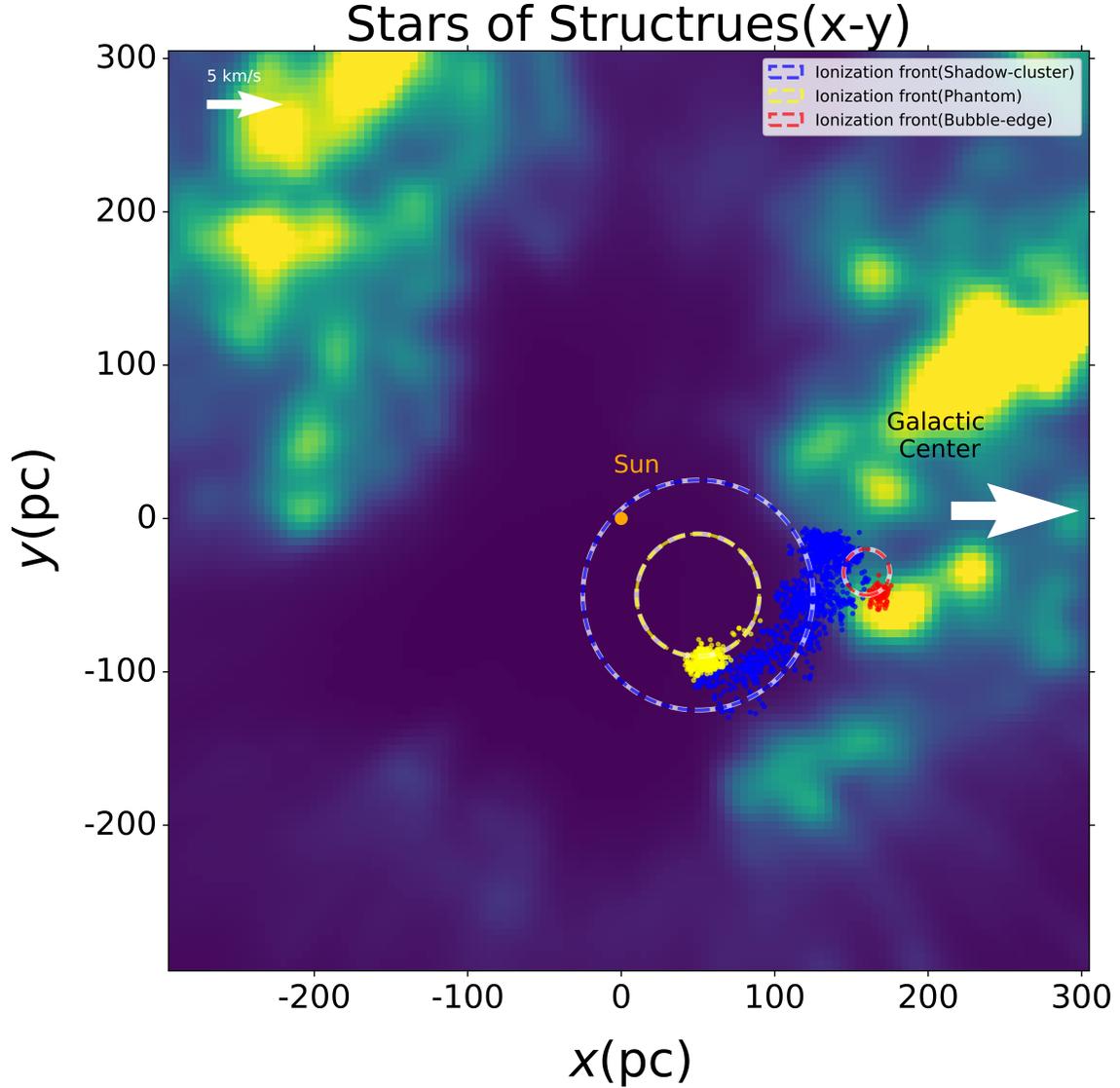}
\caption{This figure illustrates the assumptions we made during the estimation of the energy released by combined massive star action (including SN explosion). The dashed circles represent the combined massive star action (including SN explosion). We infer that the second SN explosion should have occurred between the blue dashed circle and the Bubble-edge.}
\label{fig:fig_7}
\end{figure}

To further confirm the association of these star clusters with SN explosion, we calculated the energy and momenta imparted to the star associations by the combined massive star action (including SN explosion) based on the kinematic parameters of Phantom and Shadow-cluster. As shown in Figure \ref{fig:fig_7}, we assumed that the ionization front of the SN explosion was spherical, with the explosion site located 40 pc from Phantom and 75 pc from Shadow-cluster, respectively. We used the basic kinetic energy formula and momentum formula:
\begin{equation}
    e = \frac{1}{2} m v^{2} 
\end{equation}
\begin{equation}
    p = m v
\end{equation}
\begin{equation}
    m = \rho V 
\end{equation}
\begin{equation}
    V = 4 \pi r^{2} d
\end{equation}
where the $v$ is the difference in the three-dimensional peculiar velocities of Phantom and Shadow-cluster, which we use to estimate the velocity imparted to the molecular cloud by the combined massive star action. We assume that the molecular cloud is blown to a distance of $r=65$ pc from the center by the combined massive star action, with a cloud thickness of $d=20$ pc \citep[compare][]{Krause2013}, and $\rho$ denotes the density of the ${\rm H_{2}}$ molecular cloud gas. The order of magnitude of the momentum estimation is $10^{40}$ $\rm{kg \cdot m/s}$, and the order of magnitude of the energy estimation is $10^{50}$ erg.

This order of magnitude estimate agrees well with SB simulations by \citet{Krause2013} and \citet{Krause&Diehl2014}. As Fig. 5 in \citet{Krause&Diehl2014} shows, the supershell accelerates for a few $10^{5}$ years after each supernova. The shell velocity reached 8 km/s in that particular simulation, but would be higher for slightly lower ambient density, for example. Their Fig. 2 shows the split into thermal and kinetic energy, showing that about 20\% of the supernova energy is expected to be kinetic shell energy in that phase. The shock is expected to run into the clumpy shell, temporarily increasing the density of the clumps further, thus triggering star formation \citep[][Figs 5 \& 7]{Krause2013}. We note the energetics of these simulations agree reasonably well with more recent work \citep[compare][]{Lancaster2024}.

This discovery reinforces the picture of the Local Bubble as an evolving entity and is significant for filling gaps in research on the formation and evolution of the Local Bubble.

\section{Summary} \label{sec:summary}

We study the kinematic characteristics of star associations and molecular clouds near the Local Bubble of the Sun by combining the PMS star catalog of \citet{Zari2018} with the high-precision photometry and kinematic data provided by the {\it Gaia} DR3 star catalog in order to test if the PMS associations inside the local bubble are consistent with the expectation that they have been formed in the expanding supershell. We plot the density map of the sample and extract the structures, ultimately identifying 180 structures that we certify as star associations. Finally, we calculate the kinematic data for these star associations.

We discover three unique star associations that exhibit a wiggle-like velocity pattern. The distances of Phantom, Shadow-cluster, and Bubble-edge are 108.5308, 141.5284, and 176.0318 pc, respectively, while their LSR radial velocities are 10.0621, 5.4981, and 9.0581 km/s, showing a pattern of decreasing and then increasing. We believe that they may have been influenced by multiple SN explosions during the formation of the Local Bubble, ultimately leading to this velocity pattern. We have estimated the energy released by the combined massive star action (including SN explosion), and the result's order of magnitude estimate agrees well with superbubble simulations by \citet{Krause2013} and \citet{Krause&Diehl2014}. Our research strengthens the picture of the Local Bubble as an evolving entity and is of great significance for filling the gaps in the study of the formation and evolution of the Local Bubble.

%% IMPORTANT! The old "\acknowledgment" command has be depreciated. It was
%% not robust enough to handle our new dual anonymous review requirements and
%% thus been replaced with the acknowledgment environment. If you try to 
%% compile with \acknowledgment you will get an error print to the screen
%% and in the compiled pdf.
%% 
%% Also note that the akcnowlodgment environment does not support long amounts of text. If you have a lot of people and institutions to acknowledge, do not use this command. Instead, create a new \section{Acknowledgments}.
\begin{acknowledgments}
%\section{DATA AVAILABILITY}
This work has made use of data from the European Space Agency (ESA) mission {\it Gaia} (\url{https://www.cosmos.esa.int/gaia}), processed by the {\it Gaia} Data Processing and Analysis Consortium (DPAC, \url{https://www.cosmos.esa.int/web/gaia/dpac/consortium}). Funding for the DPAC has been provided by national institutions, in particular the institutions participating in the {\it Gaia} Multilateral Agreement.
\end{acknowledgments}

\appendix

\section{The Higher LSR radial velocity} \label{APP:A}

In Section \ref{sec:SSA}, we found that the three special star associations all have LSR radial velocities higher than the typical expansion velocity of the Local Bubble. We believe the reasons should be twofold.

On the one hand, we note that the value from \citet{Zucker2022} is not a measurement of the current expansion velocity, but results from a 1D model with substantial idealisations, for example constant ambient density, energy input just from supernovae, no stellar winds, 70\% of the input energy is radiated away. We use the value therefore merely as typical value over the history of the Local Bubble.

\begin{figure}[ht!]
\plotone{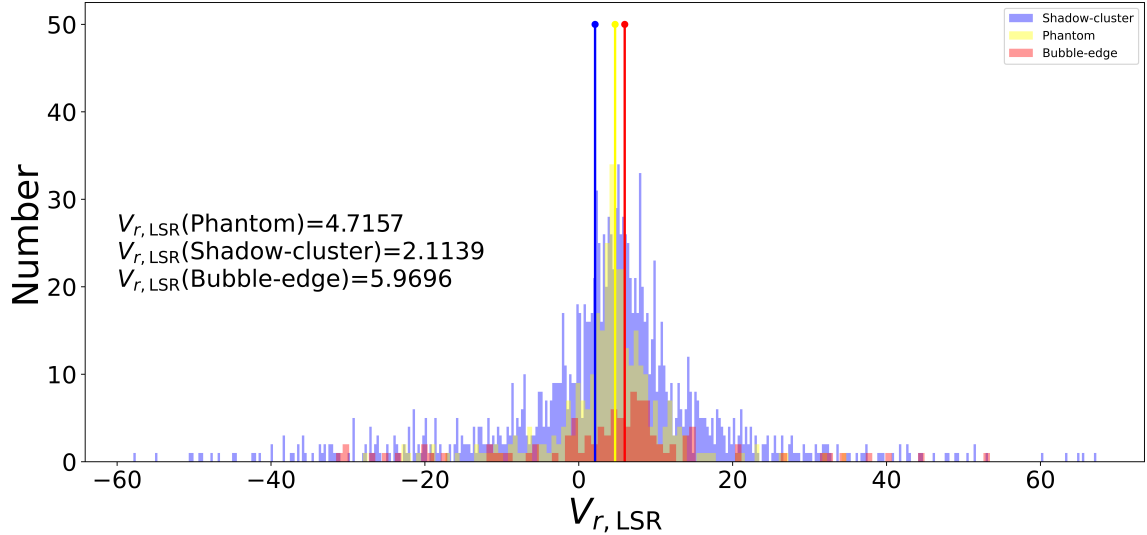}
\caption{Histogram of the LSR radial velocities of the star associations member stars calculated using parameter $U_{\odot}^{\rm{std}} = 10.0$ km/s, $V_{\odot}^{\rm{std}} = 15.4$ km/s, and $W_{\odot}^{\rm{std}} = 7.8$ km/s \citep{Kerr&Lynden-Bell1986}, with the vertical line (with a dot at the top) indicating the LSR radial velocity of the star association obtained through KDE fitting.
\label{fig:fig_8}}
\end{figure}

On the other hand, we found that the solar velocity data we used differ from those used by \citet{Zucker2022}, which are from \citet{Kerr&Lynden-Bell1986}, namely $U_{\odot}^{\rm{std}} = 10.0$ km/s, $V_{\odot}^{\rm{std}} = 15.4$ km/s, and $W_{\odot}^{\rm{std}} = 7.8$ km/s. We present the $v_{r,\rm{LSR}}$ calculated using these parameters in Figure \ref{fig:fig_8}. We found that these star associations still exhibit a wiggle-like velocity pattern in this case, but their velocities are all lower than the typical expansion velocity of the Local Bubble. The difference in parameters should be the main reason for the significant numerical discrepancy. Considering that the parameters we used are the updated solar motion parameters after Gaia, we believe that our results should be reliable.

\section{Discovery of Special Associations} \label{APP:B}

\begin{figure}[ht!]
\plotone{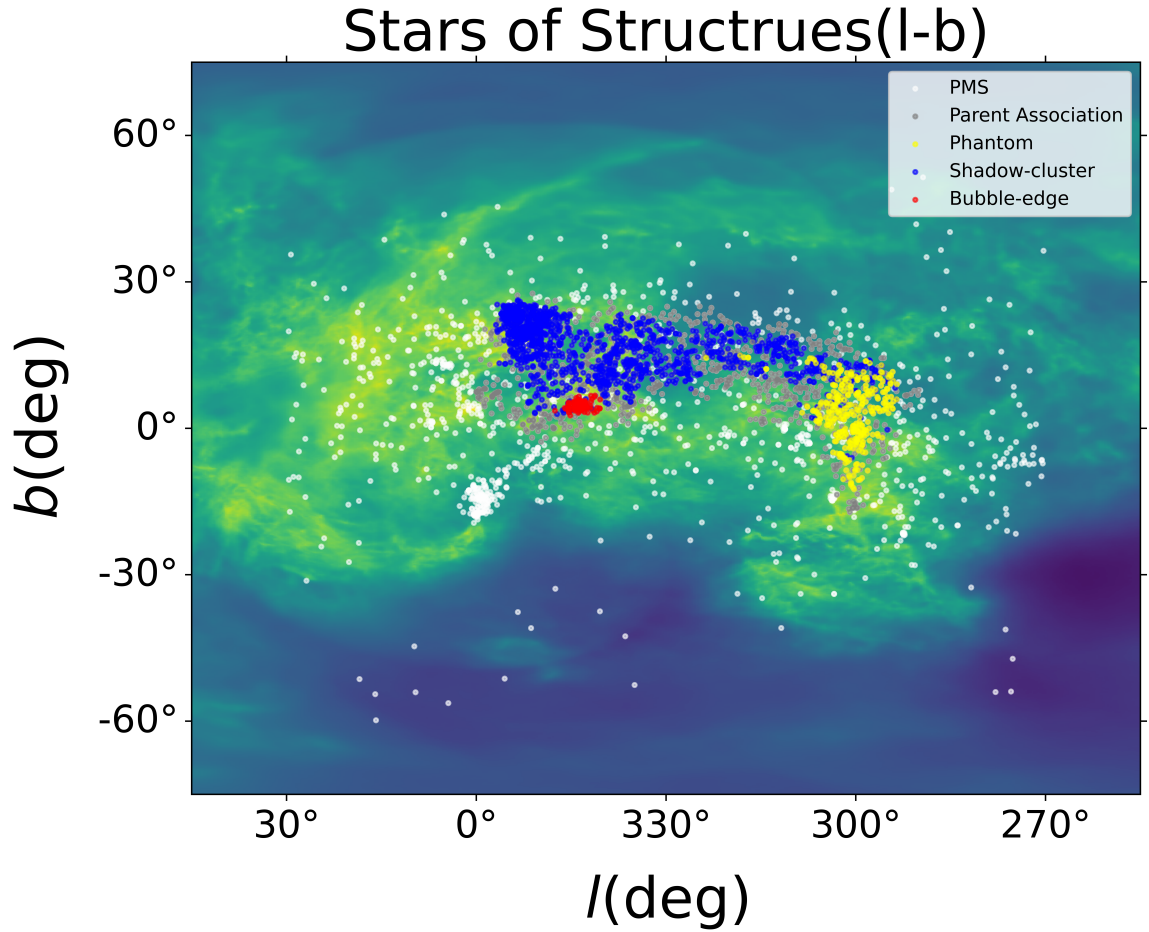}
\caption{The spatial distribution of the member stars of the parent association and the sub-associations, as well as the PMS stars in the region defined by $-90^{\circ} < l < 30^{\circ}$, $-60^{\circ} < b < 60^{\circ}$, and $80\;{\rm pc} < d < 220\;{\rm pc}$.
\label{fig:fig_9}}
\end{figure}

After calculating the kinematic parameters of the 180 star associations identified by the Dendrogram program, we inspected these star associations to identify objects of interest. In the inspection of the region defined by $-90^{\circ} < l < 30^{\circ}$, $-60^{\circ} < b < 60^{\circ}$, and $80\;{\rm pc} < d < 220\;{\rm pc}$, we focused on the most prominent series of star associations in this region. The Dendrogram program identified three sub-associations, and they all belong to the same parent association. We found that these sub-associations are spatially closely connected but exhibit a wiggle-like velocity pattern in their radial velocity distribution, which is why we are interested in them. Figure \ref{fig:fig_9} shows the spatial distribution of the member stars of the parent association and the sub-associations, as well as the PMS stars in this region.

\bibliography{sample631}{}
\bibliographystyle{aasjournal}

%% This command is needed to show the entire author+affiliation list when
%% the collaboration and author truncation commands are used.  It has to
%% go at the end of the manuscript.
%\allauthors

%% Include this line if you are using the \added, \replaced, \deleted
%% commands to see a summary list of all changes at the end of the article.
%\listofchanges

\end{document}